\documentclass[TJSASS, pubform]{tjsass} 

\usepackage[dvipdfmx]{graphicx}
\RequirePackage{multicol}
\RequirePackage{amsmath}
\RequirePackage[varg]{txfonts}
\RequirePackage{bm}
\RequirePackage{array}
\RequirePackage{color}
\RequirePackage[hang]{footmisc}
\RequirePackage{xurl}
\usepackage[colorlinks, urlcolor=blue, linkcolor=blue, citecolor=blue, plainpages=false]{hyperref}
\usepackage{upgreek}
\usepackage{placeins}

\RequirePackage{comment}

\pubyear{}
\bookvolume{}
\bookissue{}
\titleheadertrue
\receiveddate{}%
\revisiondate{}
\accepteddate{}
\confinfo{Presented at the 32$^{\text{nd}}$ International Symposium on Lepton Photon Interactions at High Energies, Madison, Wisconsin, USA, August 25-29, 2025. }
\AEname{J. Aero} 

\title{MUSIC: a detector concept for 10 TeV $\mathbf{\mu^+\mu^-}$ collisions}
\subtitle{}
\author{
\NAME{Paolo}{Andreetto},\thanksNum{1)}
\NAME{Massimo}{Casarsa},\thanksNum{2)}\CorresAuthor{massimo.casarsa@ts.infn.it}
\NAME{Alessio}{Gianelle},\thanksNum{1)}
\NAME{Donatella}{Lucchesi},\thanksNum{1,3)}\\
\NAME{Leonardo}{Palombini},\thanksNum{1,3)}
\NAME{Lorenzo}{Sestini},\thanksNum{4)}
\NAME{Davide}{Zuliani}\thanksNum{1,3)}
}
\thanksOrg{INFN - Sezione di Padova, via F. Marzolo 8, 35131 Padua, Italy}
\thanksOrg{INFN - Sezione di Trieste, via A. Valerio 2, 34127 Trieste, Italy}
\thanksOrg{Università di Padova, via F. Marzolo 8, 35131 Padua, Italy}
\thanksOrg{INFN - Sezione di Firenze, via B. Rossi 3, 50019 Sesto Fiorentino, Florence, Italy}
\begin{abstract}
The full exploitation of the physics potential of a multi-TeV muon collider will ultimately lie in the detector's ability to cope with unprecedented levels of machine-induced backgrounds. This contribution introduces the MUSIC (MUon System for Interesting Collisions) detector concept and presents its performance in the context of $\sqrt{s} = 10$~TeV muon-antimuon collisions. The MUSIC detector is designed to mitigate machine-induced background effects while maintaining high efficiency and accuracy in the reconstruction of physics events, in particular in the Higgs boson sector and in the search for new physics. It features an all-silicon tracking system, a semi-homogeneous lead-fluorite crystal electromagnetic calorimeter, a iron-scintillator sampling hadronic calorimeter, and a superconducting magnet providing a 5~T magnetic field.
Detailed detector simulations, accounting for the dominant machine-induced backgrounds, demonstrate promising performance in track, muon, photon, electron, and jet reconstruction, as well as jet flavor identification, highlighting the detector’s strong potential for high-energy muon collider experiments.
\end{abstract}
\begin{document}

\maketitle

\section{Introduction}

A muon collider represents the most appealing option for achieving leptonic collisions at multi‑TeV center‑of‑mass energies with a relatively compact circular machine~\cite{EPJC}. Such high-energy $\mu^+\mu^-$ collisions would unlock an exceptional physics program, enabling high-precision tests of the Standard Model in an unexplored energy regime, probing the structure of the Higgs sector and the shape of the Higgs potential, and allowing both direct and indirect searches for new physics.
However, the inherently unstable nature of muons introduces unique background conditions. Fully realizing the physics potential of a multi-TeV muon collider will depend critically on the detector’s capability to handle unprecedented levels of machine-induced backgrounds.

The design of a detector at a muon collider is therefore primarily driven by the physics program, constraints imposed by the machine design, and the experimental conditions along with the necessary mitigation measures~\cite{AR}.
The detector requirements driven by physics considerations are broadly aligned with those of other future multi-TeV colliders to reconstruct boosted physics objects with low transverse momenta originating from Standard Model processes, as well as central, high-energy objects produced in the decays of potential new massive states. In addition, the detector must be capable of identifying unconventional experimental signatures such as disappearing tracks, displaced leptons, photons, and jets.
On the machine side, constraints arise from the design of the interaction region and the machine-detector interface, notably the placement of final focusing quadrupoles at $\pm6$~m from the interaction point. Furthermore, the unique background conditions necessitate dedicated mitigation strategies to ensure robust detector performance and preserve sensitivity to the full breadth of the physics program.
Ultimately, the detector design, the technological choices, and the development of the event reconstruction algorithms will be driven by the high levels of machine-induced background. 

\begin{figure*}[!t]
    \centering
    \includegraphics[width=0.98\linewidth]{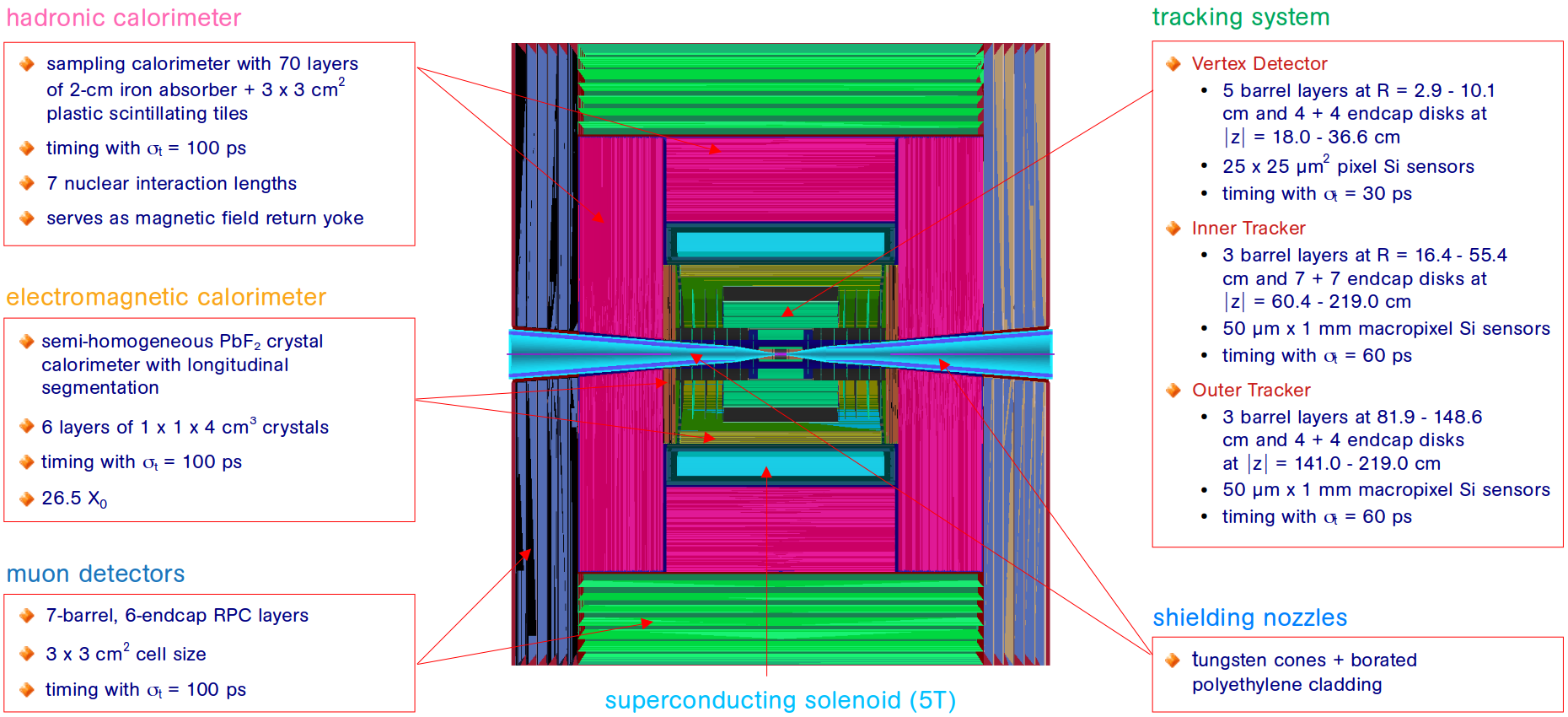}
    \caption{The MUSIC detector concept.
    \label{fig:MUSIC}}
\end{figure*}

The dominant sources of machine-induced background in the detector are the decay of muons circulating in the collider, referred to as beam-induced background, and the incoherent $e^+e^-$ pair production~\cite{daniele}.
The beam-induced background arises from high-energy electrons and positrons produced in muon beam decays, which interact with machine elements and the machine-detector interface, generating intense fluxes of low-energy secondary particles, on the order of $\mathcal{O}(10^8)$ particles per bunch crossing, in the detector. 
The resulting background is primarily composed of photons, neutrons, and electrons/positrons with energies typically below 1 GeV, a significant fraction of which reach the detector out of time relative to the bunch crossing.
In contrast, incoherent $e^+e^-$ pairs are produced through interactions of real or virtual photons emitted by the colliding beams, with an energy spectrum that peaks at a few tens of MeV and extends up to 100~GeV.
These electrons and positrons enter the detector volume at the interaction region in time with the bunch crossing, subsequently interacting with detector and machine-detector components to produce approximately $\mathcal{O}(10^6)$ secondary particles per bunch crossing, primarily photons, neutrons, and additional electrons and positrons.

\section{The MUSIC detector concept}

The MUSIC (MUon System for Interesting Collisions) detector concept has been specifically designed and optimized for $\mu^+\mu^-$ collisions at a center-of-mass energy of 10 TeV. 
It features a typical geometry for multipurpose collider experiments, with a cylindrical layout measuring 11.4 m in length and 12.8 m in diameter. Starting from the innermost region, it comprises an all-silicon tracking system followed by a calorimetric system, made up of an electromagnetic and a hadronic component. The outermost layer of the detector is occupied by a muon system.
A superconducting solenoid is placed between the electromagnetic and the hadronic calorimeters to generate a magnetic field of 5 T.
The field intensity is chosen to minimize the effects of the incoherent $e^+e^-$ pair production on the tracking system. 
Inside the detector volume along the beamline are placed two conical tungsten shields, referred to as ``nozzles'', on either side of the interaction point to screen the detector from the high-energy primary products of muon decays.
A right-handed reference system is used with the origin at the center of the detector, the nominal collision point: the $z$-axis is aligned with the direction of the clockwise-circulating $\mu^+$ beam, the $y$ axis points upward, and the $x$ axis lies on the plane of the collider ring. 
The components of the MUSIC detector are illustrated in Fig.~\ref{fig:MUSIC}. 

\FloatBarrier

\medskip
\subsubsection{Tracking system}
\smallskip

The tracking system's configuration and geometry have been optimized to maximize acceptance and provide redundant measurements that help to minimize background effects. The tracking system consists of three sub-detectors.

\smallskip
\noindent
\textbf{Vertex Detector:} It is the innermost system, consisting of five central cylindrical layers, each 26.0 cm long, positioned at radii ranging from 2.9 to 10.1 cm from the beam axis. The forward and backward regions feature four disks, oriented transverse to the beamline and located at distances of $|z| = 18.0$ to 36.6 cm from the interaction point.
A layout of $25 \times 25$ $\upmu$m$^2$ pixels is assumed with a hit spatial resolution of 5 $\upmu$m $\times$ 5 $\upmu$m and a hit time resolution of $30$ ps.

\smallskip
\noindent
\textbf{Inner Tracker:} It consists of 50 $\upmu$m $\times$ 1 mm macropixel modules arranged in three barrel layers, at radii from 16.4 to 55.4 cm, and seven disks on either side at $|z|$ from 60.4 to 219.0 cm. The first two barrel layers are 96.32 cm long, while the third measures 138.46 cm. Hit spatial and time resolutions of 7 $\upmu$m $\times$ 90 $\upmu$m and 60 ps are assumed, respectively.     

\smallskip
\noindent
\textbf{Outer Tracker:} It includes three 252.84-cm long barrel layers at radii between 81.9 and 148.6 cm and four endcap disks located at $|z|$ from 141.0 to 219.0 cm. It features 50~$\upmu$m~$\times$~1~mm macropixel modules with a hit spatial resolution of 7~$\upmu$m~$\times$~90~$\upmu$m and a hit time resolution of 60 ps.

\begin{figure*}[t]
    \centering
    \includegraphics[width=0.45\textwidth]{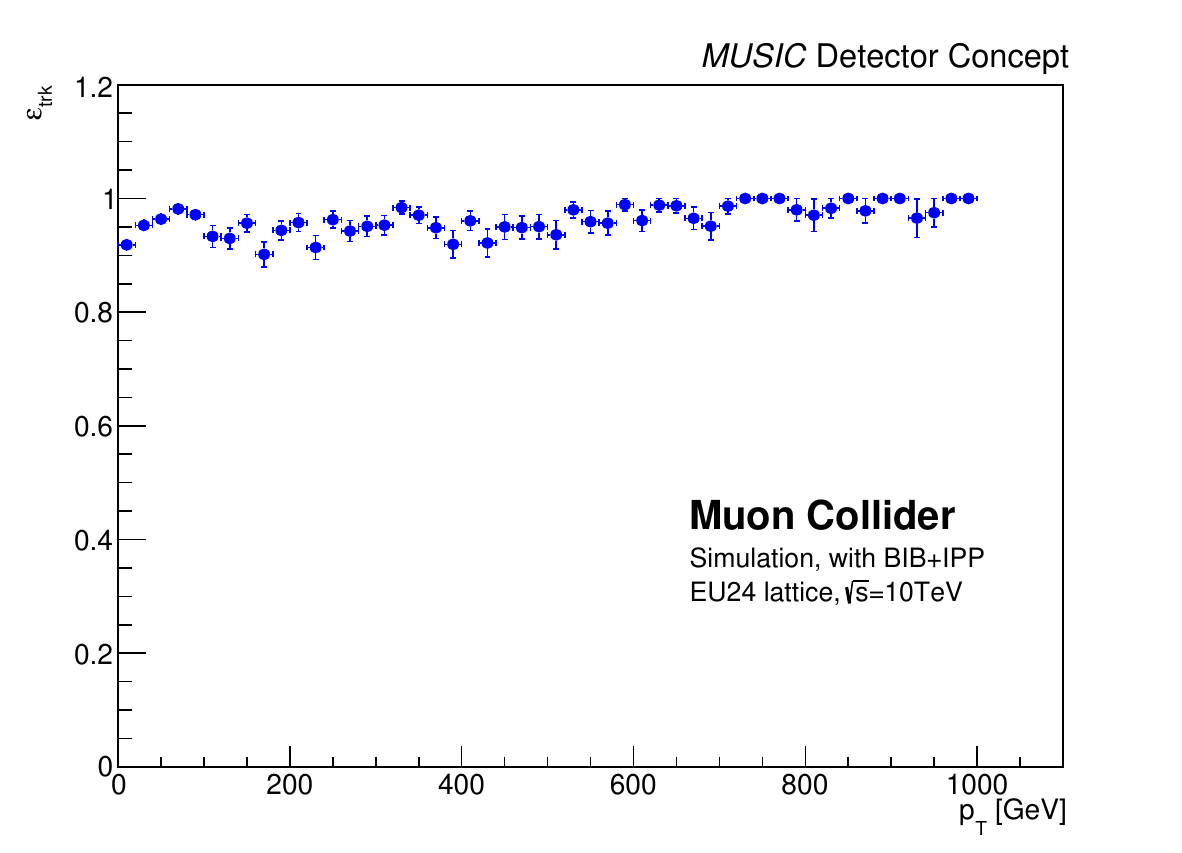} \hspace{0.5cm}
    \includegraphics[width=0.45\textwidth]{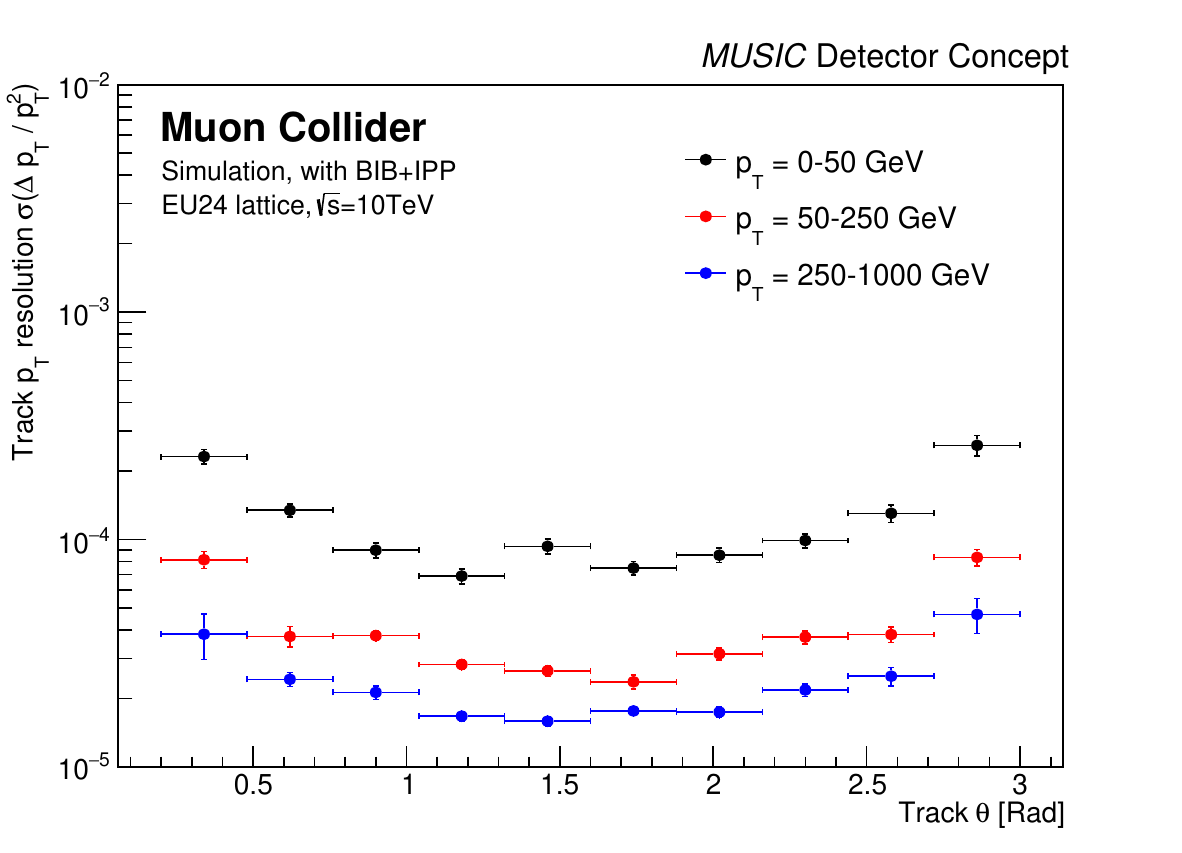}
 \caption{Left: Track reconstruction efficiency as a function of the muon transverse momentum in a sample of single muons with BIB and IPP overlaid. Right: Track transverse-momentum resolution as a function of the muon polar angle in three ranges of $p_T$. To suppress the rate of fake tracks, reconstructed from random combinations of background hits, cleaning criteria of $p_T > 1$\,GeV and $|d_0| < 0.1$\,mm are applied.}
 \label{fig:tracking}
\end{figure*}

\medskip
\subsubsection{Calorimeters}
\smallskip

The MUSIC detector comprises an electromagnetic (ECAL) and a hadronic (HCAL) calorimeter. The elec\-tro\-mag\-net\-ic calorimeter is specifically designed for muon-antimuon collisions, while the hadronic calorimeter is adapted from the CLIC Collaboration detector, which was originally designed for $e^+e^-$ collisions at 3~TeV~\cite{CLICDET}.

\smallskip
\noindent
\textbf{Electromagnetic calorimeter:} The ECAL is a semi-homogeneous crystal cal\-o\-rim\-e\-ter with longitudinal segmentation (CRILIN)~\cite{CRILIN}. It consists of $1 \times 1 \times 4$-cm$^3$ lead-fluorite crystals arranged in six layers for a total of 26.5 radiation lengths. The cylindrical barrel section has an inner radius of 169.0~cm and is 442.0~cm long. The endcaps, shaped as disks, have inner and outer radii of 31.0 cm and 196.0~cm, respectively, and are positioned at $|z| =$ 230.7~cm. 

\smallskip
\noindent
\textbf{Hadronic calorimeter:} The HCAL is an iron-scintillator sampling calorimeter composed of 70 layers of 2\,cm-thick iron absorbers interleaved with $3 \times 3$\,cm$^2$ scintillator pads, each 0.3\,cm thick, totaling approximately seven nuclear interaction lengths. It consists of a central cylindrical section measuring 501.8\,cm in length and 290.2\,cm in radius, along with two endcaps positioned at $|z| = 257.9$\,cm. The endcaps span inner and outer radii of 32.0\,cm and 475.6\,cm, respectively. Located outside the magnet, the HCAL iron absorber also serves as a return yoke for the magnetic field flux.

\medskip
\subsubsection{Muon system}
\smallskip

The final technology for the muon detectors has not yet been selected. The current detector is modeled on the re\-sis\-tive-plate chambers (RPCs) employed by the CLIC detector~\cite{CLICDET}. It features seven barrel layers, each measuring 888.8~cm in length, with radii ranging from 480.6~cm to 680.0~cm, and six endcap layers at $|z|$ positions between 444.4~cm and 590.3~cm, with inner and outer radii of 49.3 cm and 680.0~cm, respectively. The RPCs are segmented into $3 \times 3$ cm$^2$ cells. A hit time resolution $\sigma_t = 100$ ps is assumed.

\medskip
\subsubsection{Detector magnet}
\smallskip

The current configuration of the MUSIC detector includes a superconducting solenoid that generates a
uniform magnetic field of 5~T at the interaction point. The steel vacuum tank of the magnet measures
5~m in length, with an inner radius of 2~m and a total thickness of 80.7~cm.

\section{MUSIC detector performance}

The MUSIC detector concept is fully integrated into the software framework of the International Muon Collider Collaboration~\cite{MuonColliderSoft}. Its performance has been assessed through detailed detector simulations based on \textsc{Geant4}~\cite{GEANT}, incorporating machine-induced backgrounds from muon decays (beam-induced background, BIB) and incoherent $e^+e^-$ pair production (IPP).

Detector performance studies rely on high-statistics samples of both BIB and IPP. The BIB samples are generated using the FLUKA simulation toolkit~\cite{FLUKA}, which includes a detailed model of the machine lattice at the interaction region and the machine-detector interface. These components are specifically designed for 10~TeV collisions and optimized to suppress backgrounds reaching the detector, in a configuration referred to as the ``EU24 lattice''~\cite{daniele}. 
The IPP background is produced using the Guinea-Pig generator~\cite{guinea-pig}, with subsequent propagation and interactions of the resulting electrons and positrons with the nozzles simulated with FLUKA.

The following sections present reconstruction efficiencies and parameter resolutions for tracks, muons, photons, electrons, and hadronic jets, evaluated using detailed detector simulations of single-particle and dijet samples, with BIB and IPP overlaid on an event-by-event basis.
Detector performance was further assessed through full-fledged physics analyses aimed at estimating the sensitivity, given the current detector configuration and reconstruction algorithms, on the Higgs boson production cross sections for the channels $H \to b\bar{b}$, $H\to WW^\ast$, and $H\!H \to b\bar{b}b\bar{b}$, as well as on the Higgs boson trilinear self-coupling, using a detailed detector simulation with BIB and IPP overlaid on both signal and physics background events. Details are provided in Ref.~\cite{Leonardo}.

\medskip
\subsubsection{Tracks}
\smallskip

\begin{figure*}[t]
 \centering
    \includegraphics[width=0.45\textwidth]{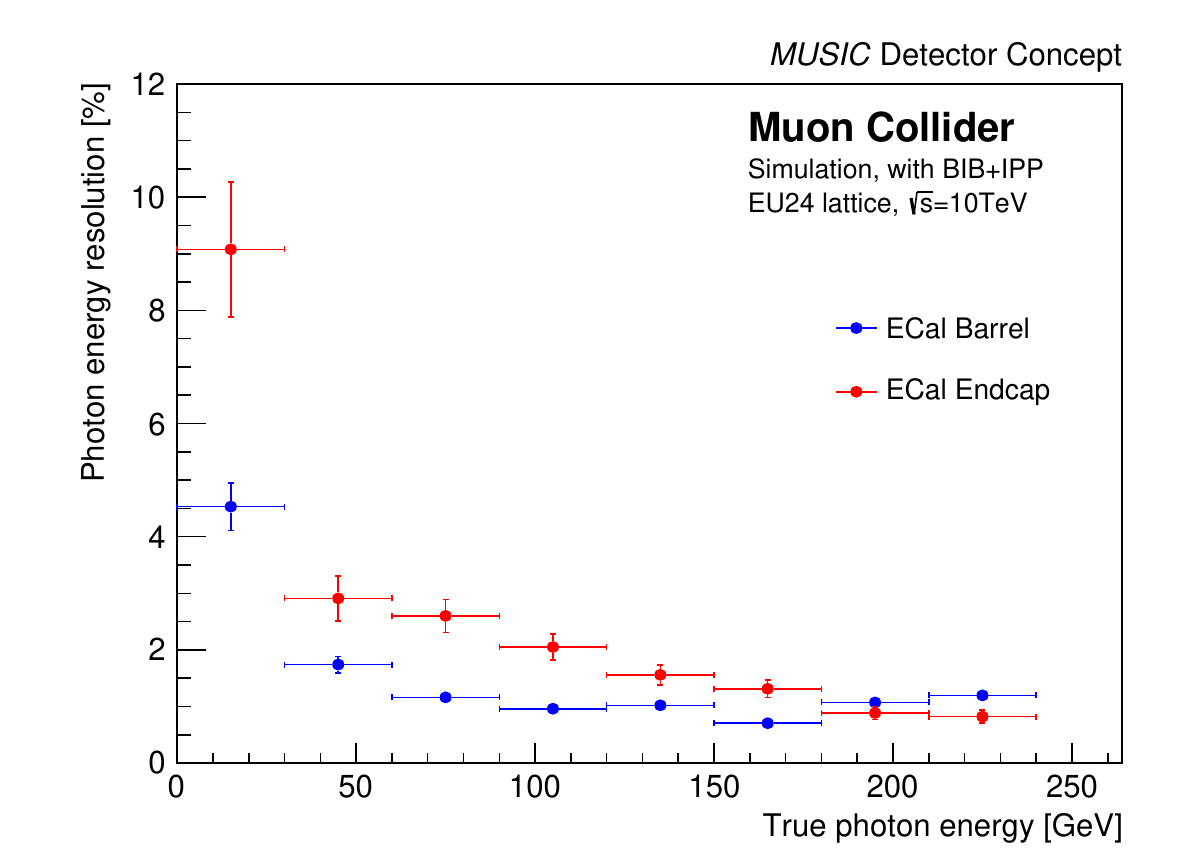}\hspace{0.5cm}
    \includegraphics[width=0.45\textwidth]{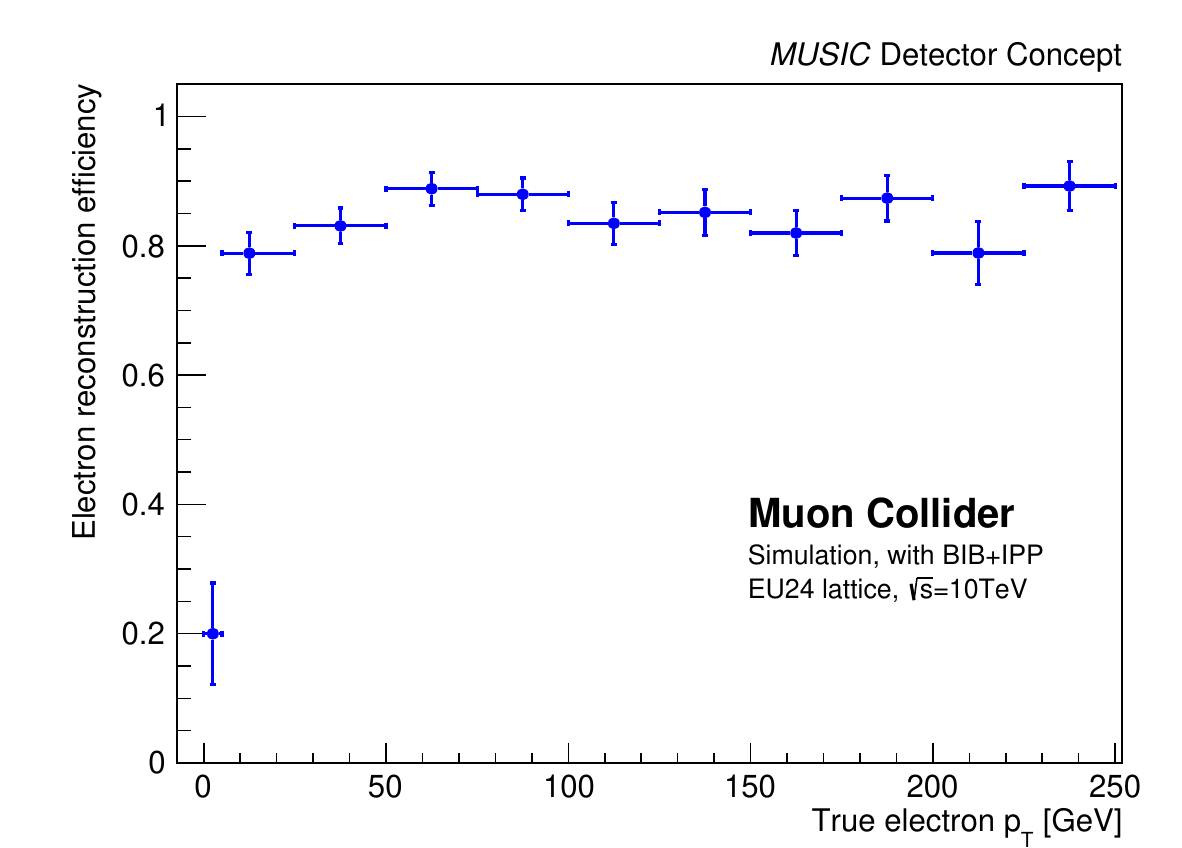}
 \caption{Left: Photon energy resolution in ECAL barrel and endcaps as a function of photon energy in a single photon sample with BIB and IPP overlaid. Right: Electron reconstruction efficiency in the ECAL barrel as a function of electron $p_T$ in a single electron sample with BIB and IPP overlaid.
 \label{fig:photons}}
\end{figure*}

Being the closest subdetector to the beamline, the tracking system is the most affected by machine-induced backgrounds, particularly in the innermost layer of the vertex-detector barrel, where an average density of approximately 5000 spurious hits/cm$^2$ is expected within a readout window of $[-0.5, 15]$\,ns relative to the bunch crossing. To mitigate this, tighter timing constraints are applied, selecting only hits whose timestamps are consistent with particles originating from the interaction point, within a window of $[-3\sigma_t, +5\sigma_t]$, where $\sigma_t$ denotes the hit time resolution.

Tracks are then reconstructed using a Combinatorial Kalman Filter implemented within the ACTS package~\cite{ACTS}. Seeding is performed using hits from the second, third, and fourth layers of the vertex detector. The first layer is excluded from this step due to its high hit occupancy caused by machine-induced background, which would otherwise result in an unmanageable number of combinatorial seed candidates.
Nonetheless, the first layer is incorporated during the track fitting stage to ensure high precision in impact parameter determination.

Tracking performance has been studied using samples of single muons, with beam-induced and incoherent $e^+e^-$ pair backgrounds overlaid on an event-by-event basis. As shown in the left panel of Fig.~\ref{fig:tracking}, the tracking reconstruction efficiency exceeds 95\% over a broad range of muon transverse momenta $p_T$, where $p_T$ denotes the component of momentum perpendicular to the beamline. The resolution $\Delta p_T/p_T^2$, illustrated in the right panel of Fig.~\ref{fig:tracking}, reaches a few $10^{-5}$ for $p_T > 50$\,GeV in the central region of the tracking system, and gradually deteriorates at lower momenta and at polar angles closer to the beamline.

\medskip
\subsubsection{Muons}
\smallskip

Muons are identified by matching tracks to hits in the outer muon system. Performance, evaluated using single prompt muons overlaid with BIB and IPP, shows an identification efficiency above 90\% for $p_T > 5$\,GeV, stable across polar angles with a slight drop of a few percent at barrel-endcap transitions. On average, one fake muon per event with $p_T > 5$\,GeV is expected from background.

\medskip
\subsubsection{Photons and electrons}
\smallskip

The ECAL is affected by a diffuse, approximately uniform machine-induced background, primarily composed of soft photons. In the first layer of the ECAL barrel, an average energy density of about 400\,MeV/cm$^2$ is expected, while the energy deposition in the endcaps is roughly an order of magnitude lower. To mitigate the effects of this background, hit energy thresholds, based on both hit arrival time and energy, have been optimized across different ECAL layers and polar angle regions.
ECAL hits are grouped into clusters using particle-flow algorithms provided by the PandoraPFA package~\cite{Pandora}. Subsequently, energy corrections are applied to the clusters as a function of their energy and polar angle, in order to compensate for reconstruction inefficiencies and detector-related effects.

Photon candidates are defined as ECAL clusters without corresponding tracks, consistent with the expectation for neutral particles.
The photon reconstruction efficiency approaches 100\% within the angular range $20^\circ < \theta_\gamma < 160^\circ$, but drops to approximately 60\% in the endcap regions, around $\theta \simeq 10^\circ$ and $170^\circ$. As shown in the left panel of Fig.~\ref{fig:photons}, the estimated energy resolution is $\Delta E/E \simeq 10\%/\sqrt{E\,\text{[GeV]}}$ in the barrel and $\Delta E/E \simeq 17\%/\sqrt{E\,\text{[GeV]}}$ in the endcaps. The reduced performance in the endcaps is attributed to a higher flux of background photons, which is expected to be mitigated through an optimized nozzle design as well as through more advanced reconstruction algorithms.

Electrons are identified as electromagnetic clusters matched to reconstructed tracks. The overall identification efficiency, driven by both track reconstruction and track-cluster matching, is approximately 85\% in the barrel region for $p_T > 10$\,GeV, as shown in Fig.~\ref{fig:photons} (right). The energy resolution for electrons in the ECAL barrel is comparable to that of photons. In the endcap regions, however, electron identification remains challenging due to the higher density of fake tracks in the forward direction, which limits performance and is the focus of ongoing optimization efforts.

\medskip
\subsubsection{Hadronic jets}
\smallskip

\begin{figure*}[t]
 \centering
    \includegraphics[width=0.45\textwidth]{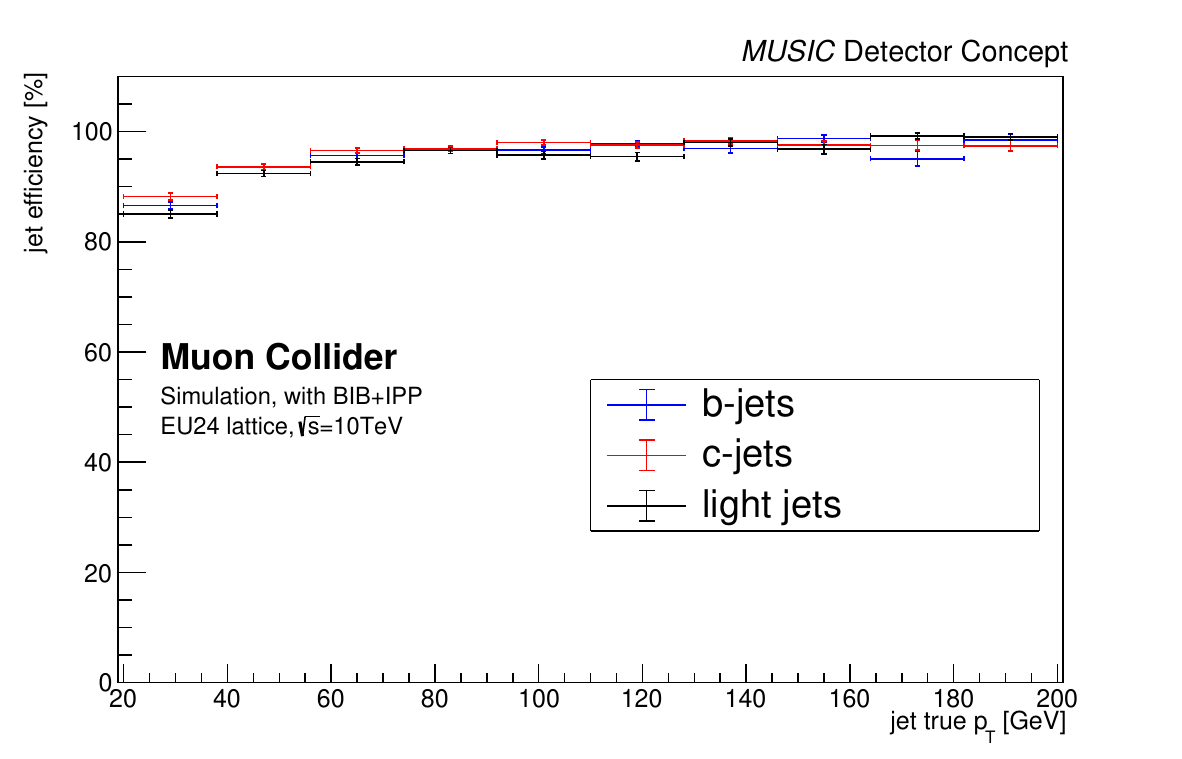}
    \includegraphics[width=0.45\textwidth]{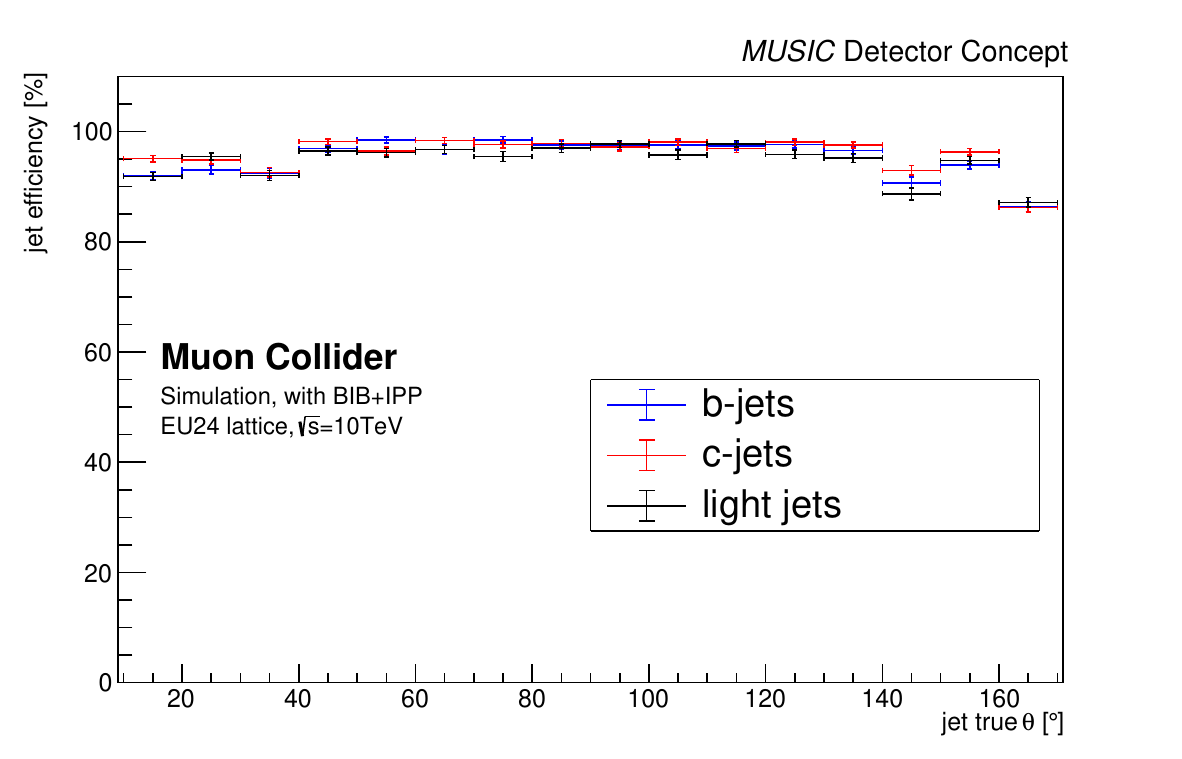} \\
    \hspace*{-1.15cm}\includegraphics[width=0.45\textwidth]{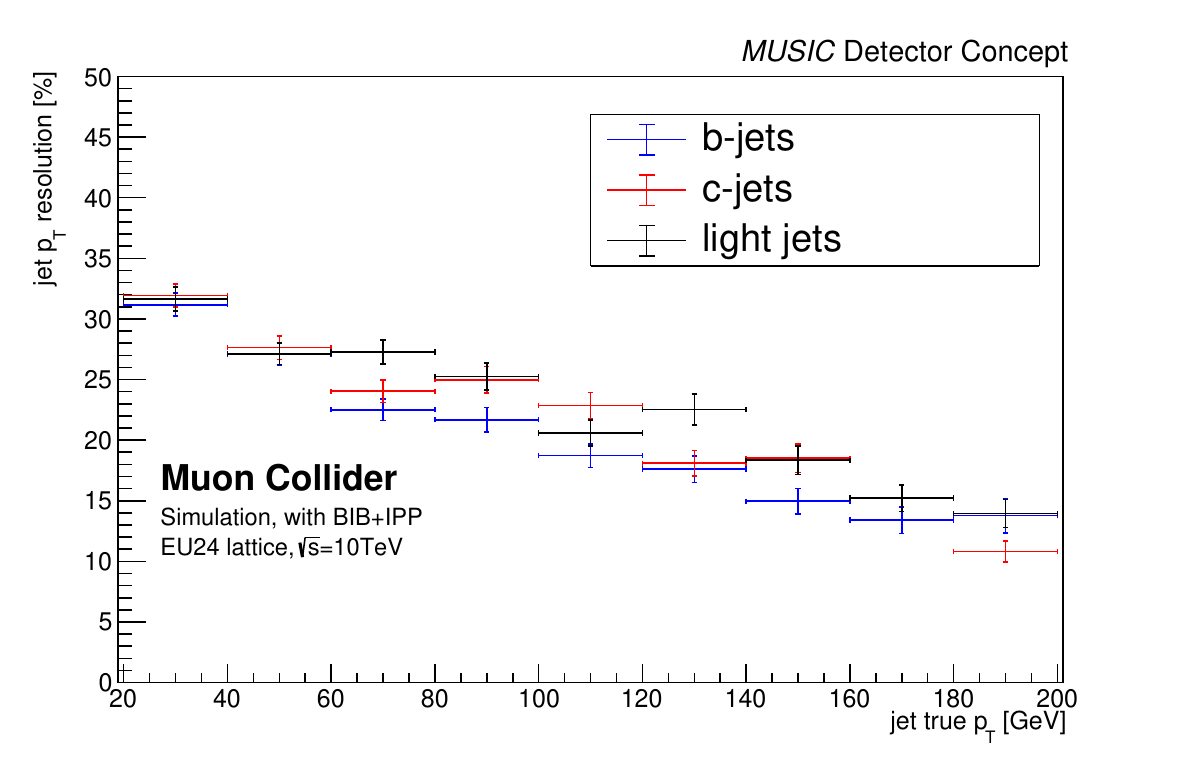}
    \hspace{0.7cm}\includegraphics[width=0.345\textwidth]{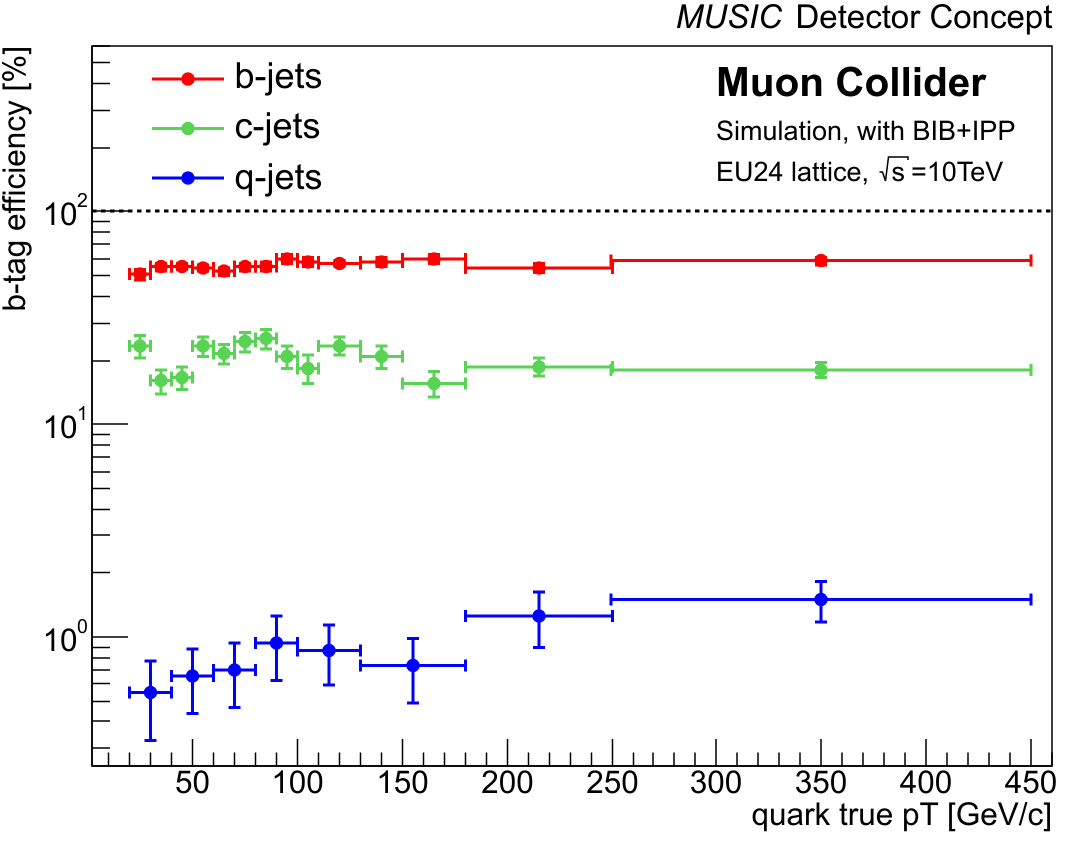}
    \caption{Top: Reconstruction efficiency as a function of generator-level jet $p_T$ (left) and polar angle (right) for jets originating from $b$-, $c$-, and light-flavor quarks. 
    Bottom left: Jet energy resolution as a function of generator-level jet $p_T$ for $b$-, $c$-, and light-flavor jets in the central detector region ($60^\circ < \theta_{\text{jet}} < 120^\circ$). 
    Bottom right: $b$-tagging efficiency as a function of the true quark $p_T$ for jets originating from $b$-, $c$-, and light-flavor quarks.
    \label{fig:jets}}
\end{figure*}

Hadronic jets originate from the hadronization of color-charged partons. Experimentally, they are complex objects whose reconstruction relies on information from all subdetectors.
Jet reconstruction begins with tracks and calorimeter clusters that, after applying minimal cleaning criteria, serve as inputs to a particle-flow algorithm based on the PandoraPFA software~\cite{Pandora}. This algorithm has been specifically optimized for the muon collider environment and reconstructs individual charged and neutral particles by combining information from the tracking and calorimeter systems. The resulting particle-flow objects are then clustered into jets using the $k_t$ algorithm~\cite{kt_algo} with a radius parameter of $R = 0.5$, chosen to optimize the dijet mass resolution and reconstruction efficiency for Higgs boson decays to $b\bar{b}$ final states. The jet four-momentum is calculated by summing the four-momenta of its constituent particles. 
To account for detector effects and reconstruction biases, a correction, derived from comparisons between reconstructed and generator-level jets, is applied to jet $p_T$ and, in the barrel-endcap transition regions, to jet direction.

Identifying jets from $b$ quarks ($b$ tagging) is crucial for many physics analyses, including those targeting $H \to b\bar{b}$ decays. The $b$ tagging algorithm exploits distinctive signatures of $b$-hadron decays, such as displaced secondary vertices within the jet and the presence of displaced muons. These features are combined using two Deep Neural Networks: one trained to discriminate \textit{b}-jets from light-quark jets, and the other to separate \textit{b}-jets from \textit{c}-jets.

Jet reconstruction and $b$-tagging performance have been evaluated using samples of $b\bar{b}$, $c\bar{c}$, and light-quark dijet events, generated with Pythia~\cite{PYTHIA} with an approximately flat distribution in dijet invariant mass. These events were fully simulated and reconstructed, including machine-induced background. To suppress fake jets from background, a selection requiring at least one track per jet with a maximum $p_T$ above 2\,GeV is applied, resulting in an average of 0.85 fake jets per event.
Jet reconstruction efficiencies as a function of jet $p_T$ and polar angle are shown in the top panels of Fig.~\ref{fig:jets} for different jet flavors. The efficiency reaches approximately 85\% at $p_T \sim 20$\,GeV and exceeds 95\% for $p_T > 60$\,GeV. In the central region, the efficiency approaches 100\% and remains stable for all jet flavors.
The $p_T$ resolution for central jets, shown in Fig.~\ref{fig:jets} (bottom left), is about 30\% at $p_T \sim 20$\,GeV and improves to 10\% at $p_T \sim 200$\,GeV. At low $p_T$, the resolution degrades to approximately 45\% in the barrel-endcap transition regions and in the endcaps.
Figure~\ref{fig:jets} (bottom right) shows the $b$-tagging efficiency as a function of the true quark $p_T$ for jets originating from $b$-, $c$-, and light-flavor quarks. The results correspond to a representative working point, yielding an average efficiency of 55\% for \textit{b}-jets, 20\% for \textit{c}-jets, and 0.8\% for light-flavor jets.

\section{Conclusion}

A new detector concept, MUSIC, has been developed specifically for 10 TeV $\mu^+\mu^-$ collisions. 
MUSIC has been extensively employed in detector performance studies and full-fledged Higgs boson analyses that are based on a detailed detector simulation including the dominant machine-induced backgrounds from muon decays and incoherent $e^+e^-$ pair production.
The results demonstrate strong reconstruction performance for key physics objects and competitive sensitivity to Higgs sector measurements, even under challenging background conditions, highlighting the detector’s significant potential for high-energy muon collider experiments.

\section*{Acknowledgments}\label{Acknowledgments}
{\small 
This work was supported by the European Union’s Horizon 2020 and Horizon Europe Research and
Innovation programs through the Marie Sk\l{}odowska-Curie RISE Grant Agreement No. 101006726
and the Research Infrastructures INFRADEV Grant Agreement No. 101094300.
}


\end{document}